\documentclass[reqno,10pt]{amsart}
\usepackage{xspace}
\newcommand{\bd}{\begin{definition}}                
\newcommand{\ed}{\end{definition}}                  
\newcommand{\bc}{\begin{corollary}}                 
\newcommand{\ec}{\end{corollary}}                   
\newcommand{\bl}{\begin{lemma}}                     
\newcommand{\el}{\end{lemma}}                       
\newcommand{\bp}{\begin{proposition}}            
\newcommand{\ep}{\end{proposition}}                
\newcommand{\bere}{\begin{remark}}                  
\newcommand{\ere}{\end{remark}}                     

\def\beq#1\eeq{\begin{equation}#1\end{equation}}     
\def\bal#1\eal{\begin{align}#1\end{align}}          %
\def\baln#1\ealn{\begin{align*}#1\end{align*}}      
\def\bml#1\eml{\begin{multline}#1\end{multline}}    %
\def\bmln#1\emln{\begin{multline*}#1\end{multline*}}
\def\bga#1\ega{\begin{gather}#1\end{gather}}
\def\bgan#1\egan{\begin{gather*}#1\end{gather*}}    
\def\bsp#1\esp{\begin{split}#1\end{split}}
\DeclareMathOperator{\de}{d}                        
\newcommand{\N}{\ensuremath{\mathbb{N}}\xspace}     
\newcommand{\R}{\ensuremath{\mathbb{R}}\xspace}     
\newcommand{\To}{\longrightarrow}                   
\newcommand{\abs}[1]{\ensuremath%
            {\vert#1\vert}\xspace}                     
\newcommand{\inte}{\int_0^1\!\!}
\newcommand{\ptl}{\ensuremath{\omega}\xspace}
\newcommand{\eh}{\ensuremath{F}\xspace}

\newcommand{\dd}{{\rm d}}
\newcommand{\p}{\partial}

\newcommand{\gkk}{\ensuremath{g^{\mathrm{kk}}}}
\newtheorem{theorem}{Theorem}[section]
\newtheorem{corollary}[theorem]{Corollary}
\newtheorem{lemma}[theorem]{Lemma}
\newtheorem{proposition}[theorem]{Proposition}
\theoremstyle{definition}
\newtheorem{definition}[theorem]{Definition}
\theoremstyle{remark}
\newtheorem{remark}[theorem]{Remark}

\begin{document}

\title[Solutions to the Lorentz force equation]{Solutions to the Lorentz force equation  with fixed
charge-to-mass ratio in globally hyperbolic spacetimes}

\author[E. Caponio]{Erasmo Caponio}
\address{Dipartimento Interuniversitario di Matematica,
Universit\`a - Politecnico di Bari, via E. Orabona 4, 70125, Bari,
ITALY} \email{caponio@dm.uniba.it}

\author[E. Minguzzi]{Ettore Minguzzi}
\address{Dipartimento di Fisica,
Universit\`a di Milano-Bicocca, Piazza della Scienza 3, 20126
Milano, ITALY} \email{minguzzi@mib.infn.it}




\begin{abstract}
We extend the classical Avez-Seifert theorem  to trajectories of
charged test particles with fixed charge-to-mass ratio. In
particular, given two events $x_{0}$ and $x_{1}$, with $x_{1}$ in
the chronological future of $x_{0}$, we find an interval
$I=]-R,R[$ such that for any $q/m \in I$ there is a timelike
connecting solution of the Lorentz force equation. Moreover, under
the assumption that there is no null geodesic connecting $x_0$ and
$x_1$, we prove that to any value of $|q/m|$ there correspond at
least two connecting timelike solutions which coincide only if
they are geodesics.
\end{abstract}
\maketitle
\section{Introduction}
Let $\Lambda$ be a  Lorentzian manifold endowed with the metric
$g$ having signature $(+ - - -)$, and consider a  point particle
of rest mass $m$ and electric charge $q$, moving in the
electromagnetic field $\eh$.

The equation of motion is the so called {\em Lorentz force
equation} (cf. \cite{mtw}) \beq D_s \left(\frac{\de x}{\de
s}\right)=\frac{q}{mc^2}\hat F(x)\left[\frac{\de x}{\de s}\right],
                                                                                \label{lorentzeq}
\eeq where $x=x(s)$ is the world line of the particle, $\frac{\de
x}{\de s}$ is its four-velocity, $D_s\left(\frac{\de x}{\de
s}\right)$ is the covariant derivative of $\frac{\de x}{\de s}$
along $x(s)$ associated to the Levi-Civita connection of $g$, and
$\hat F(x)[\cdot]$ is the linear map  on $T_x \Lambda$ metrically
equivalent to $F(x)$,  that is
\[
g(x)[v,\hat F(x)[w]] = F(x)[v,w],
\]
for any $v,\ w\in T_x\Lambda$.

We state the following question:\par
\medskip
\noindent {\em Does equation \eqref{lorentzeq} have at least one
timelike future-oriented  solution connecting two given events
$x_{0}$ and $x_{1}$ with $x_{1}$ in the chronological future of
$x_{0}$, for any charge-to-mass ratio $q/m$?}\par
\medskip

It is well known (see for instance \cite{BEE}) that, if the
manifold $\Lambda$ is globally hyperbolic, the Avez-Seifert
theorem gives a positive answer to the above question in the case
$q=0$ (i.e. for the {\em geodesic equation}).

In this paper we prove that, for an exact electromagnetic field on
a  globally hyperbolic manifold $\Lambda$,  the answer is positive
for any ratio $q/m$ in  a suitable neighborhood of $0\in\R$.

Our strategy is to derive the solutions to the Lorentz force
equations as projections of geodesics of a higher dimensional
manifold. In this way we are able to use the techniques already
developed for the geodesic equation.

This approach,  has been already  used for studying the Lorentz
force equation  in General Relativity  (see for instance \cite{K})
but a result, {\em a la} Avez and Seifert, for  the Lorentz force
equation was still lacking.

In the preprint \cite{CM}  it was proved that in a globally
hyperbolic spacetime  Eq. \eqref{lorentzeq} admits a connecting
solution with a charge-to-mass ratio different from zero. That
ratio, however was not fixed since the beginning.

So assume that $\eh$ is an exact two-form and let \ptl be a
potential one-form for \eh. Let us consider a trivial bundle $P=\Lambda \times
\mathbb{R}$, $\pi : P \to \Lambda$, with the structure group $T_{1}:$ $ b \in T_{1}$,
$p=(x,y)$, $p'=p b=(x, y+b)$, and $\tilde\ptl$ the connection one-form on $P$:
\[
\tilde\ptl=i(\dd y + \frac{e}{\hbar c} \ptl).
\]
Here  $y$ is a dimensionless coordinate on the fibre, $-e \,
(e>0)$ is the electron charge and $\hbar=h/2\pi$, with $h$ the
Planck constant. Henceforth we will denote by $\bar \ptl$ and
$\bar\eh$, respectively the one-form $\frac{e}{\hbar c} \ptl$ and
the two-form $\frac{e}{\hbar c} \eh$. Let us endow $P$ with the
Kaluza-Klein metric
\begin{equation}
\gkk=g+a^{2}  \tilde\omega^{2}  \label{kk}
\end{equation}
or  equivalently, using the notation $z$ for the points in $P$ and
the identification $z=(x,y)\in \Lambda\times\R$,
\[
\gkk(z)[w,w]=\gkk(x,y)[(v,u),(v,u)]=g(x)[v,v]- a^{2}(u +
\bar\ptl(x)[v])^{2},
\]
for every $w=(v,u)\in T_x \Lambda\times \R$. The positive constant
$a$ has the dimension of a length and  has been introduced for
dimensional consistency of definition \eqref{kk}. In the
compactified five-dimensional Kaluza-Klein theory the fibre is
isomorphic to $S^1$ and  $a$ represents the radius of the fifth
dimension.

Let us consider the Lagrangian on $P$ \baln L=L(z,w):
TP\To\R&&L(z,w)=\frac{1}{2} \gkk(z)[w,w]. \ealn Fix two points
$p_{0}$ and $p_{1} \in P$. The geodesics on $P$, with respect to
the Kaluza-Klein metric, connecting the points $p_0$ and $p_1$ are
the critical points of the action functional
\[
S=S(z)=\inte\frac{1}{2} \gkk(z(\lambda))[\dot z(\lambda),\dot
z(\lambda)]\de \lambda,
\]
defined  on  a suitable space of sufficiently regular curves on
$P$, parameterized from $0$ to $1$, with fixed extreme points
$p_0$ and $p_1$. Here $\dot z$ denotes the derivative of
$z=z(\lambda)$ with respect to $\lambda$.

Assume that $z(\lambda)=(x(\lambda),y(\lambda))$ is a critical
point for $S$. Since the Lagrangian $L$ is independent of $y$, the
following quantity $p_z$ is conserved
\[
p_{z}=\frac{\p L}{\p \dot{y}}=-a^{2}(\dot{y} +
\bar\ptl(x)[\dot{x}]).
\]
Moreover taking variations only with respect to  the variable $x$
we obtain the following equation   for $x=x(\lambda)$
\beq
\label{x} D_{\lambda} \dot x=p_{z} \hat{\bar{F}}(x)[\dot x].
\eeq
From \eqref{x}, it follows that $g(x)[\dot x,\dot x]$ is
constant along $x$. Assume that $x$ is non-spacelike (with respect
to $g$) and define $C \ge 0$ such that
\[
g(x)[\dot x,\dot x] =C^{2}.
\]
Since $z$ is a geodesic, also  $\gkk(z)[\dot z,\dot z]$ is
conserved and
\begin{equation}\label{gkkconst}
\gkk(z)[\dot z,\dot z]=C^{2}-\frac{p_{z}^{2}}{a^{2}}.
\end{equation}
Thus the geodesic $z$ on $P$ is timelike iff
\[
C^{2} > \frac{p_{z}^{2}}{a^{2}}.
\]
\bere\label{null} Of course, if $z$ is timelike then also $x$ is
timelike, and if $z$ is non-spacelike then also $x$ is non-spacelike. Moreover, if $z$ is
a null geodesic, then $C^{2}  =  p_{z}^{2}/a^{2}$ and $x$ is timelike iff
$p_z\neq 0$.
\ere

\bere\label{comparison} Now assume that $x$ is timelike. The
proper time for $x$ is defined by
\[
\dd s=C \dd \lambda,
\]
hence if we parameterize $x$ with respect to proper time, from
\eqref{x}, we get the following equation for $x=x(s)$
\[
D_s \left(\frac{\de x}{\de s}\right) =
  \frac{p_z}{C}\hat{\bar F}(x)\left[\frac{\de x}{\de s}\right] =
     \frac{p_z}{C}\frac{e}{\hbar c}\hat F(x)\left[\frac{\de x}{\de s}\right].
\]
Therefore, a comparison with \eqref{lorentzeq} allows us to
conclude that, if we are able to find a future-oriented null
geodesic for the Kaluza-Klein metric, starting from a point
$p_0=(x_0,y_0)$, arriving to a point $p_1=(x_1,y_1)$ and having
constant $p_z\neq 0$, then, recalling Remark~\ref{null}, we can
state that there exists  a future-oriented timelike solution to
\eqref{lorentzeq}, connecting $x_0$ and $x_1$ and having
charge-to-mass ratio
\[
\frac{q}{m}=\frac{p_z}{C}\frac{ec}{\hbar}=\pm\frac{aec}{\hbar},
\]
with the plus sign if $p_z>0$ and the minus sign if $p_z<0$. \ere

\section{Statement and proof of the main theorem}
In this section we state and prove our main result. In the sequel
we will make large use of the notations of the book \cite{HE},
which is our reference also for the necessary background on causal
techniques.

Let  $\mathcal{T}_{x_0,x_1}$  and $\mathcal{N}_{x_0,x_1}$ be the
sets, respectively, of all the  $C^1$, future-pointing timelike
connecting curves and of all the $C^1$, future-pointing
non-spacelike connecting curves. With {\em connecting curve} we
mean a map $x$ from an interval $[a,b]\subset \R$ to $\Lambda$
such that $x(a)=x_{0}$ and $x(b)=x_{1}$ and any other map $w$ such
that $w=x\circ\lambda$ with $\lambda$ a
 $C^1$ function from an interval
$[c,d]$ to the interval $[a,b]$, having positive derivative.

Define
\begin{equation}
R=\sup_{x \in \mathcal{T}_{x_0,x_1}} \Big(\frac{c^{2} \int_{x} \dd
s}{ \sup_{w \in \mathcal{N}_{x_0,x_1}}|\int_{w} \omega - \int_{x}
\omega|}\Big). \label{L}
\end{equation}
Notice  that $R$ does not depend on the gauge chosen, that is, it
is invariant under the replacement $\omega \to \omega+ \eta$ where
$\eta$ is an exact one-form.

We recall that a globally hyperbolic manifold is a Lorentzian
manifold  containing  a subset (a so called {\em Cauchy surface})
which is intersected by every inextendible non-spacelike smooth
curve precisely once.

\bp\label{Mfinite} Let $(\Lambda,g)$ be a time-oriented, globally
hyperbolic, Lorentzian manifold,  let $x_0$ be a point on
$\Lambda$ and $x_1\in\Lambda$ a point in the chronological future
of $x_0$. Then
\[R>0.\]
\ep
\begin{proof}
Let $\gamma$ be a connecting future-directed timelike geodesic
whose length is
\[
L=\sup_{x \in \mathcal{N}_{x_0,x_1}} \int \de s.
\]
For the Avez-Seifert theorem, such a geodesic  exists. Choose a
gauge such that $\int_{\gamma} \omega=0$, and define
\[
M=\sup_{x \in \mathcal{N}_{x_0,x_1}} | \int_{x} \omega |.
\]
We are going to prove that $M$ is finite. Let
$\{x_n\}_{n\in\N}\subset\mathcal N_{x_0,x_1}$ be a sequence such
that
\[ |\int_{x_n}\omega| \To M.\]
Since $(\Lambda,g)$ is globally hyperbolic, the set $C(x_0,x_1)$
of continuous non-spacelike  curves connecting $x_0$ and $x_1$ is
compact \cite{HE}. We recall that the topology of $C(x_0,x_1)$ is
defined by saying that a neighborhood of $\eta \in C(x_0,x_1)$
consists of all the curves in $C(x_0,x_1)$ whose points in
$\Lambda$ lie in a  neighborhood $W$ of the points of $\eta \in
\Lambda$.  We can extract a subsequence, denoted again with
$\{x_n\}$, such that $x_n$ converges  to a continuous
non-spacelike curve $x$ on $\Lambda$, connecting $x_0$ and $x_1$
in the topology on $C(x_0,x_1)$. Since $x$ is compact we can cover
it with $m$ charts $(U_k,\phi_k)$ of the form
\begin{equation}
\phi_{k}: U_k \to \Delta^{4} \subset \mathbb{R}^{4} \quad
\textrm{with} \ \Delta=]0,b[,
\end{equation}
where the coordinates $\{ x^{\mu}_{k} \}$ are Gaussian normal
coordinates
\begin{equation}
g_{\mu \nu} \de x^{\mu} \de x^{\nu}= (\de x^{0}_{k})^{2} -
\gamma_{i j \, k}(x^{0}_k, x^{i}_k) \de x^{i}_{k} \de x^{j}_{k},
\end{equation}
and $\partial_{0 k}$ is future-directed (the existence of a
neighborhood of $p \in M$ having Gaussian coordinates  follows by
lemma 4.5.2 of \cite{HE}). Here $\gamma_{i j \, k}$ is a positive
definite metric on the spacelike hypersurfaces of constant
$x^{0}_k$.  Moreover we can assume that $x_0\in U_1$, $x_1\in
U_m$, and $U_i\cap U_{k}=\emptyset$, for any $k\neq i-1, i, i+1$.
 Let us introduce in $U_k$, the inverse of $\gamma_{ij \, k}$,
$\gamma^{-1 ij}_{k}$, and the function
$\tilde{\omega}^{i}_{k}=\gamma^{-1 ij}_{k} \omega_{j \, k}$ where
$\omega_{j \, k}$ denote the components of $\ptl$ in $U_k$.  In
$U_{k}$ we consider the continuous functions $\omega_{0k}$ and
$\sqrt{\tilde{\omega}^{i}_{k} \gamma_{ij \, k}
\tilde{\omega}^{j}_{k}}$. Since $x$ is compact, we can find a
neighborhood $W \subset \bigcup_{1 \le k \le m} U_{k}$ of $x$, and
a  constant $C$, such that for any  $k$, $|\omega_{0 k}|<C$ and
$\sqrt{\tilde{\omega}^{i}_{k} \gamma_{ij \, k}
\tilde{\omega}^{j}_{k}}<C$. Since $x_n$ converges to $x$ there is
an  integer number $N$ such that, for $n>N$, $x_n \in W$.

Moreover,  for any $x_n$, $n >N$, the strong causality condition on $\Lambda$
allows us to   introduce  a partition
$\{[\lambda_{k-1}, \lambda_k]\}_{1\leq k\leq m}$,
$\lambda_0=0<\lambda_1<\ldots<\lambda_m=1$, of the interval
$[0,1]$, such that
  $x_n([\lambda_{k-1},\lambda_k])\subset
U_k \cap W$. Now we compute
\baln
|\int_{x_n} \omega| &\le
\int_{0}^{1} | \omega(x_n)[\dot{x}_{n}] | \de \lambda =
\sum_{k=1}^{m} \int_{\lambda_{k-1}}^{\lambda_{k}}
 | \omega(x_n)[\dot{x}_{n}] | \de \lambda \\
&= \sum_{k=1}^{m}
\int_{\lambda_{k-1}}^{\lambda_{k}}  | \omega_{0k} \dot{x}^{0}_{n}
+\omega_{ik} \dot{x}^{i}_{n} | \de \lambda
=\sum_{k=1}^{m} \int_{\lambda_{k-1}}^{\lambda_{k}}  | \omega_{0k}
\dot{x}^{0}_{n} +\tilde{\omega}^{i}_{k} \gamma_{i j \, k}
\dot{x}^{j}_{n} | \de \lambda.
\ealn

Using the Schwarz inequality we have
\begin{equation}\label{schwarz}
|\tilde{\omega}^{i}_{k} \gamma_{i j \, k} \dot{x}^{j}_{n}| \le
\sqrt{(\tilde{\omega}^{i}_{k} \gamma_{i j \, k}
\tilde{\omega}^{j}_{k})( \dot{x}^{s}_{n} \gamma_{s l \, k}
\dot{x}^{l}_{n}) } \le C  \sqrt{\dot{x}^{i}_{n} \gamma_{i j\, k}
\dot{x}^{j}_{n}} \le C \dot{x}^{0}_{n}
\end{equation}
where in the last step we have used the fact that $x_{n}$ in
non-spacelike and future-directed.  From \eqref{schwarz} we get
\begin{eqnarray*}
|\int_{x_n} \omega| \le \sum_{k=1}^{m}
\int_{\lambda_{k-1}}^{\lambda_{k}}  2C \dot{x}^{0}_{n}  \de
\lambda \le 2Cmb < +\infty
\end{eqnarray*}
Passing to the limit on $n$, we conclude. Finally, recalling the
definition of $R$, we get
\[
R \ge \frac{c^{2} L}{M} > 0.
\]

\end{proof}
Now we are ready to state our main result.
\begin{theorem}\label{main}
Let $(\Lambda,g)$ be a time-oriented Lorentzian manifold. Let \ptl
be a one-form ($C^{2}$) on $\Lambda$ (an electromagnetic
potential) and $\eh=\de \ptl$ (the electromagnetic tensor field).
Assume that $(\Lambda,g)$ is a globally  hyperbolic manifold. Let
$x_{1}$ be an event in the chronological future of $x_{0}$ and let
$R$  be defined as in \eqref{L}, then there exists at least one
future-oriented timelike solution to \eqref{lorentzeq} connecting
$x_0$ and $x_1$, for any charge-to-mass ratio satisfying
\begin{equation}
\abs{\frac{q}{m}}< R  \label{qsum}
\end{equation}
\end{theorem}

Before proving Theorem~\ref{main} we need some lemmas. The first
is the following result, about the causal structure of the
manifold $P$, which is contained in \cite{CM}. We report the proof
for the reader convenience.

\bl The manifold $P=\Lambda \times \mathbb{R}$ endowed with the
metric \eqref{kk} is a time-oriented globally hyperbolic
Lorentzian manifold. \el
\begin{proof}
 Let $V$ be a timelike vector field on $\Lambda$ giving a time orientation.
Clearly the horizontal lift of $V$, $(V, -\bar\omega[V])$, gives a
time-orientation to $P$ (henceforth we will consider $P$
time-oriented by means of such  a vector field).

Let us prove that if $z\colon]a, b[\to P$ ($-\infty\leq
a<b\leq+\infty$) is an inextendible smooth future-pointing
non-spacelike curve, then $x(\lambda)=\pi(z(\lambda))$ is an
inextendible smooth future-pointing non-spacelike curve.
By contradiction let $o$ be a future endpoint for $x$
corresponding to $s=b$.  We are going to prove that $z$ has a
future endpoint $u$, $\pi(u)=o$. Since $z$ is non-spacelike we
deduce that
\[ a\abs{\dot y+\bar\omega(x)[\dot x]}\leq
        \sqrt{g(x)[\dot x,\dot x]},\]
and, integrating from $c>a$ to $d<b$, we get \beq
a\int_c^d\abs{\dot y+\bar\omega(x)[\dot x]}\de \lambda\leq
    \int_c^d\sqrt{g(x)[\dot x,\dot x]}\de \lambda.
    \label{fromctod}
\eeq
Now consider the Lorentzian distance function $d$ on $\Lambda$
associated to the metric $g$. Since $\Lambda$ is globally
hyperbolic and  $x$ is non-spacelike, the right-hand side of
\eqref{fromctod} is less than $d(x(c),x(d))<+\infty$. As $x$ has
future endpoint $o$ corresponding to $\lambda=b$,
$d(z(c),o)<+\infty$. So  there exists the limit as $d\to b^-$ of
the right-hand side of \eqref{fromctod}. Therefore  the left-hand
side of  \eqref{fromctod} has finite limit as $d\to b^-$. Now
consider the term $\int_c^d\bar\omega(x)[\dot x]\de \lambda$. Pick
a Gaussian coordinate system $(U_o,\varphi)$ at $o$ as in the
proof of Proposition~\ref{Mfinite}. Without loss of generality we
can assume that $x(c)\in U_o$ and $x(d)\in U_o$, for any $c\leq
d\leq b$. Denote $x(\lambda)$ by $(x^0(\lambda),x^i(\lambda))$
for any $\lambda\in[c,b]$. Since $x$ is non-spacelike, we have
\beq
 \gamma_{ij}(x^0,x^i)\dot x^i\dot x^j\leq(\dot x^0)^2.
                    \label{causal2}
\eeq Moreover as $x$ is future-pointing,  $\dot x^0(\lambda)\neq
0$ on $[c,b[$, thus $x^0(\lambda)$ is strictly monotone on
$[c,b[$.

Arguing as in the proof of Proposition~\ref{Mfinite}, we obtain
\baln
\int_c^d\abs{\bar\omega(x)[\dot x]}\de \lambda&
=\int_c^d\abs{\bar\omega_0\dot x^0+\bar\omega_i\dot x^i}\de \lambda\\
&=\int_c^d\abs{\bar\omega_0\dot x^0+\bar{\tilde\omega}^i\gamma_{ij}\dot x^j}\de
\lambda\leq\int_c^d2C\dot x^0\de \lambda.
\ealn

Passing to the limit as $d\to b^-$, we conclude that
$\abs{\omega(x)[\dot x]}$ is integrable on $[c, b]$. As
\[\lim_{d\to b^-}\int _c^d(\dot y+\bar\omega(x)[\dot x])\de \lambda\in\R,\]
we conclude that
\[
\lim_{d\to b^-}y(d)-y(c)=\lim_{d\to b^-}\int_c^d \dot y\de
\lambda\in\R.
\]
Let $\bar y=\lim_{d\to b^-}y(d)$. Clearly the point $(o,\bar y)\in
P$ is a future endpoint for $z$ corresponding to $s=b$. This fact
yields the desired contradiction.

Now, let $S$ be a Cauchy
surface for $\Lambda$, then $\tilde S=S\times\R$ is a
Cauchy surface for $P$. Indeed $z(\lambda)$ meets $\tilde S$ as
many times as $x(\lambda)$ meets $S$, and in correspondence of the
same value of the parameter. Since $S$ is a Cauchy surface for
$\Lambda$, $x(\lambda)$ meets $S$ exactly once and $z(\lambda)$ meets
$\tilde S$ exactly once.
\end{proof}
\bere\label{E+} Let $E^{+}(p_{0})=J^{+}(p_{0})-I^{+}(p_{0})$,
$p_{0} \in P$. It is well known (see  \cite[p. 112,184]{HE}) that
if $q \in E^{+}(p_{0})$ there exists a null geodesic connecting
$p_{0}$ and $q$. \ere \bl\label{causallysimple} Any globally
hyperbolic Lorentzian manifold $\Lambda$ is causally simple, i.e.
for every compact subset $K$ of $\Lambda$,
$\dot{J}^{+}(K)=E^{+}(K)$,  where $\dot{J}^{+}(K)$ denotes the
boundary of ${J}^{+}(K)$.
\el
\begin{proof}
See \cite[p. 188, 207]{HE}.
\end{proof}
\bl \label{con} Let $p_{0}=(x_{0},y_{0})$ and
$p_{1}=(x_{1},y_{1})$ be two points in $P$. Let us denote by
$\delta$ the difference $y_1-y_0$. Moreover let $\sigma$ be a
connecting future-oriented timelike curve. If
\begin{equation} \label{timel}
|\delta+\int_{\sigma} \bar\omega|<\frac{\int_{\sigma}\de s}{a},
\end{equation}
then  $p_1$ belongs to the chronological future of $p_0$.
\el
\begin{proof}
Let $\lambda$ be the affine parameter of $\sigma$ such that $\de s/
\de \lambda=C=\int_{\sigma} \de s$ and
 $\sigma(\lambda)_{|\lambda=0}=x_0$.
Consider the curve on $P$
\[
\tau = \tau(\lambda) =\Big(\sigma(\lambda), y_{0}+
(\delta+\int_{\sigma}\bar\omega)\lambda-\int_{0}^{\lambda}
\bar\omega[\dot{\sigma}] \de \lambda'\Big).
\]
Clearly $\tau(0)=p_0$, $\tau(1)=p_1$ and the following quantity is
constant over $\tau$
\[
\dot{y}+\bar\omega[\dot{\sigma}]=\delta+\int_{\sigma}\bar\omega.
\]
Moreover,
\[
    \gkk(\tau)[\dot \tau,\dot \tau] = C^{2}-a^{2}(\delta+\int_{\sigma}\bar\omega)^{2}.
\]
Thus, if \eqref{timel} holds, $\tau$ is a timelike future-oriented
curve that connects $p_{0}$ and $p_{1}$.
\end{proof}
\begin{proof}[Proof of Theorem~\ref{main}]
Let $\bar{R}= \frac{
\hbar}{ec} R$. In Eq. \eqref{kk} choose $a < \bar{R}$. There is a
connecting timelike curve $\sigma$ such that
\begin{equation} \label{pass}
\sup_{w \in \mathcal{N}_{x_0,x_1}}|\int_{w} \bar\omega -
\int_{\sigma} \bar\omega| < \frac{\int_{\sigma} \de s}{a}.
\end{equation}
Consider its horizontal lift $\sigma^{*}$ having initial point
$p_{0}=(x_{0}, y_{0})$. Since $\sigma^{*}$ is timelike, its final
point $\tilde{p}_{1}=(x_{1}, \tilde{y}_{1})=(x_{1},
y_{0}-\int_{\sigma} \bar{\omega})$ belongs to $I^{+}(p_{0})$. Let
$U$ be the open subset of $\R$ containing all the values $y_{1}$
such that $p_1=(x_{1}, y_{1})$ is in the chronological future of
$p_{0}$. Moreover let $V$ be the connected component of $U$
containing $\tilde y_1$. Assume that $V$ is given by
$]\bar{y}_{1},\hat{y}_{1}[$. We are going to show that
$\bar{y}_{1}>-\infty$ and $\hat{y}_{1}<+\infty$.  By
contradiction, assume that for any $y_{1} > \tilde{y}_{1}$ ($y_{1}
< \tilde{y}_{1}$), it is $p_{1}=(x_{1}, y_{1}) \in I^{+}(p_{0})$.
For the Avez-Seifert theorem there is a timelike future-oriented
geodesic $\alpha(\lambda)=(x(\lambda), y(\lambda))$ that connects
$p_{0}$ to $p_{1}$. Here $\lambda$ is the affine parameter such
that $\alpha(0)=p_{0}$ and $\alpha(1)=p_{1}$. Then there exist
constants $C_{\alpha}$ and $p_{\alpha}$ such that
$g(x)[\dot{x},\dot{x}]=C_{\alpha}^{2}$ and
$p_{\alpha}=-a^{2}(\dot{y}+\bar\omega)$.

Integrating the last equation from $0$ to $1$ gives
\[p_{\alpha}=-a^{2}(\delta+\int_{x}\bar\omega),\]
and, recalling \eqref{gkkconst},
\[
\gkk(\alpha)[\dot\alpha,\dot\alpha]=C_{\alpha}^{2}-a^{2}(\delta+\int_{x}\bar\omega)^{2},
\]
but we know (see the proof of Proposition~\ref{Mfinite}) that
$\sup_{x \in \mathcal{N}_{x_0,x_1}} |\int_{x}\bar\omega|=B<
+\infty$. For $|\delta|>L/a+B$, we obtain that $\alpha$ is
spacelike and thus a contradiction.

Now we  consider the points in $P$, $\bar{p}_{1}=(x_{1},
\bar{y}_{1})$ and $\hat{p}_{1}=(x_{1}, \hat{y}_{1})$.  By
Remark~\ref{E+} and Lemma \ref{causallysimple} there exist  two
null geodesics $\bar\eta=(\bar x,\bar y)$ and $\hat \eta=(\hat
x,\hat y)$ connecting $p_{0}$ to $\bar p_1$ and $\hat{p}_{1}$,
respectively. By Remark~\ref{null} we know that if
$p_{\bar\eta},p_{\hat\eta}\neq 0$, then the non-spacelike curves
$\bar x$ and $\hat x$  are actually timelike. Since both
$\bar{p}_{1}$ and $\hat{p}_{1}$ are in $\dot{I}^{+}(p_{0})$, from
lemma \ref{con} and \eqref{pass}, we have
\[
\sup_{w \in \mathcal{N}_{x_0,x_1}}|\int_{w} \bar\omega -
\int_{\sigma}
\bar\omega|<|\hat{y}_{1}-y_{0}+\int_{\sigma}\bar\omega|,
\]
and analogously for  $\bar{y}_{1}$. In particular
\[
|\int_{\hat{x}} \bar\omega - \int_{\sigma}
\bar\omega|<|\hat{y}_{1}-y_{0}+\int_{\sigma}\bar\omega|,
\]
and analogously with the hat replaced with a bar. Recalling that
$\tilde y_1=y_0- \int_{\sigma}\bar\omega$ and $\hat{y}_{1}
>\tilde{y}_{1}$ and $\bar{y}_{1} < \tilde{y}_{1}$, we have
\[
p_{\hat\eta}=-a^{2}(\hat{y}_{1}-y_{0}+ \int_{\hat{x}} \bar\ptl)<0,
\]
and
\[
p_{\bar\eta}=-a^{2}(\bar{y}_{1}-y_{0}+\int_{\bar{x}} \bar\ptl)>0.
\]
Therefore we have proved that there exist two timelike
future-oriented connecting solutions to the equation
\eqref{lorentzeq} having charge-to-mass ratios
\[
\frac{q}{m}=- \frac{a e c}{\hbar},
\]
and
\[
\frac{q}{m}=+ \frac{a e c}{\hbar}.
\]
Since $a<\bar{R}$ is arbitrary we get the thesis.
\end{proof}

\begin{theorem}\label{2}
Let $(\Lambda,g)$ be a time-oriented Lorentzian manifold. Let \ptl
be a one-form ($C^{2}$) on $\Lambda$ (an electromagnetic
potential) and $\eh=\de \ptl$ (the electromagnetic tensor field).
Assume that $(\Lambda,g)$ is a globally  hyperbolic manifold. Let
$x_{1}$ be an event in the chronological future of $x_{0}$ and
suppose there is no null geodesic connecting $x_{0}$ and $x_{1}$,
then there exist at least two future-oriented timelike solutions
to  Eq. \eqref{lorentzeq} connecting $x_0$ and $x_1$, for any
given value of $|q/m|$. The two curves coincide only if they are
geodesics.
\end{theorem}
\begin{proof}
Take an arbitrary timelike connecting curve $\sigma$ and consider
its horizontal lift $\sigma^{*}$. From $\sigma^{*}$, the steps of
the previous proof led to two null  geodesics over $P$. Repeating
those steps here, it follows the existence of non-spacelike
connecting curves $\hat x$ and $\bar x$ satisfying Eq. \eqref{x}.
By Remark~\ref{null}, the constants of the motion $p_{\hat \eta}$
and $p_{\bar \eta}$ do not vanish, otherwise the curves $\hat x$
and $\bar x$ would be null geodesics connecting $x_0$ and $x_1$.
By Remark~\ref{comparison}, $\hat x$ and $\bar x$, parameterized
by proper time, are timelike future oriented solutions of Eq.
\eqref{lorentzeq} having charge-to-mass ratio $q/m$ satisfying
\[
|\frac{q}{m}|=\frac{a e c}{\hbar}
\]
Let them coincide, and  denote them by $x=\hat x=\bar x$, then
$\int_{\hat x} \de s=\int_{\bar x} \de s=C$. Moreover
\[
p_{\hat\eta}-p_{\bar\eta}=-a^{2}(\hat{y}_{1}-\bar{y}_{1}) \ne 0
\]
Therefore, subtracting the Lorentz force equations satisfied by
both $\bar{x}$ and $\hat{x}$, we get
\begin{equation}
(p_{\hat\eta}-p_{\bar\eta})\hat F(x)\left[\frac{\de x}{\de
s}\right] =0  \ \Rightarrow \ \hat F(x)\left[\frac{\de x}{\de
s}\right]=0.
\end{equation}
Substituting back this equation into the Lorentz force equation we
see that $x$ is a geodesic. Since $a$ is arbitrary we obtain the
thesis.
\end{proof}

\bc Let $(M,\eta)$ be the Minkowski spacetime. Let \ptl be a
one-form on $M$  and $\eh=\de \ptl$ an electromagnetic tensor
field.  Let $x_{1}$ be an event in the chronological future of
$x_{0}$, then there exist at least two future-oriented timelike
solutions to \eqref{lorentzeq} connecting $x_0$ and $x_1$, for any
given value of $|q/m|$. The two curves coincide only if they are
geodesics. \ec
\begin{proof}
It follows from the fact that, in Minkowski spacetime, if $x_{1}
\in I^{+}(x_{0})$ there is no null geodesic connecting $x_{0}$
with $x_{1}$.
\end{proof}

\section{Conlusions}
From a physical point of view Eq. \eqref{L}  shows that for sufficiently  weak fields
$F$,  Theorem \ref{main} answers affirmatively to the existence of connecting
future-oriented timelike solutions of the Lorentz force equation. Indeed, under the
replacement $\omega \to k \omega$, $R$ scales as $R \to R/k$.
Moreover the electron is the free particle with the maximum value
of the charge-to-mass ratio, and for sufficiently small $k$,
$\frac{e}{m_e} < R$.

In case the electromagnetic field is not weak, in order to have
$q/m \ge R$, Eq. \eqref{L} shows that the electromagnetic
``energy" $q \omega[\frac{\de x}{\de s}]$ should be of the same
order of the rest energy $m c^{2}$. In this case quantum effects
may become relevant and in particular the effect of pair creation.
Theorem \ref{2} shows that, if a pair is created at the event
$x_0$, then at least one of the two particles has the ability,
with a suitable impulse, to reach the event $x_{1}$. Notice that
here we are neglecting the reciprocal electromagnetic interaction
between the particles. In a strong electromagnetic field this is,
however, allowed.

We conclude than in a classical regime the problem of the
existence of timelike connecting solutions to the Lorentz force
equations is solved. It remains open the problem of the existence
of solution in a strong field $F$, i.e. in a quantum mechanical
regime. Under these conditions we have given a partial result that
can be useful when studying the consequences of the pair creation
effect.

\end{document}